\documentclass[pra,twocolumn,aps,showpacs,preprintnumbers,amsmath,amssymb,superscriptaddress]{revtex4-1}

\usepackage{graphicx}
\usepackage{dcolumn}
\usepackage{bm}
\usepackage{amsmath}
\usepackage{hyperref}
\usepackage{braket}
\usepackage{multirow}

\begin{document}

\title{Chip-Based Measurement-Device-Independent Quantum Key Distribution}

\author{Henry Semenenko}
\email{henry.semenenko@bristol.ac.uk}
\affiliation{Quantum Engineering Centre for Doctoral Training, H. H. Wills Physics Laboratory \& Department of Electrical and Electronic Engineering, University of Bristol, Tyndall Avenue, BS8 1FD, UK}
\affiliation{Quantum Engineering Technology Labs, H. H. Wills Physics Laboratory \& Department of Electrical and Electronic Engineering, University of Bristol, Tyndall Avenue, Bristol, BS8 1FD, UK}

\author{Philip Sibson}
\affiliation{KETS Quantum Security, Unit DX, St Philips Central, Albert Road, St. Philips, Bristol, BS2 0XJ, UK}

\author{Andy Hart}
\affiliation{Quantum Engineering Technology Labs, H. H. Wills Physics Laboratory \& Department of Electrical and Electronic Engineering, University of Bristol, Tyndall Avenue, Bristol, BS8 1FD, UK}

\author{Mark G. Thompson}
\affiliation{Quantum Engineering Technology Labs, H. H. Wills Physics Laboratory \& Department of Electrical and Electronic Engineering, University of Bristol, Tyndall Avenue, Bristol, BS8 1FD, UK}

\author{John G. Rarity}
\affiliation{Quantum Engineering Technology Labs, H. H. Wills Physics Laboratory \& Department of Electrical and Electronic Engineering, University of Bristol, Tyndall Avenue, Bristol, BS8 1FD, UK}

\author{Chris Erven}
\affiliation{Quantum Engineering Technology Labs, H. H. Wills Physics Laboratory \& Department of Electrical and Electronic Engineering, University of Bristol, Tyndall Avenue, Bristol, BS8 1FD, UK}
\affiliation{KETS Quantum Security, Unit DX, St Philips Central, Albert Road, St. Philips, Bristol, BS2 0XJ, UK}

\date{\today}

\begin{abstract}
Modern communication strives towards provably secure systems which can be widely deployed. Quantum key distribution provides a methodology to verify the integrity and security of a key exchange based on physical laws. However, physical systems often fall short of theoretical models meaning they can be compromised through uncharacterised side-channels. The complexity of detection means that the measurement system is a vulnerable target for an adversary. Here we present secure key exchange up to $200$ km while removing all side-channels from the measurement system. We use mass-manufacturable, monolithically integrated transmitters that represent an accessible, quantum-ready communication platform. This work demonstrates a network topology that allows secure equipment sharing which is accessible with a cost-effective transmitter, significantly reducing the barrier for widespread uptake of quantum-secured communication. 
\end{abstract}

\maketitle

\section{Introduction}

\begin{figure*}
	\centering
	\includegraphics[width = \linewidth]{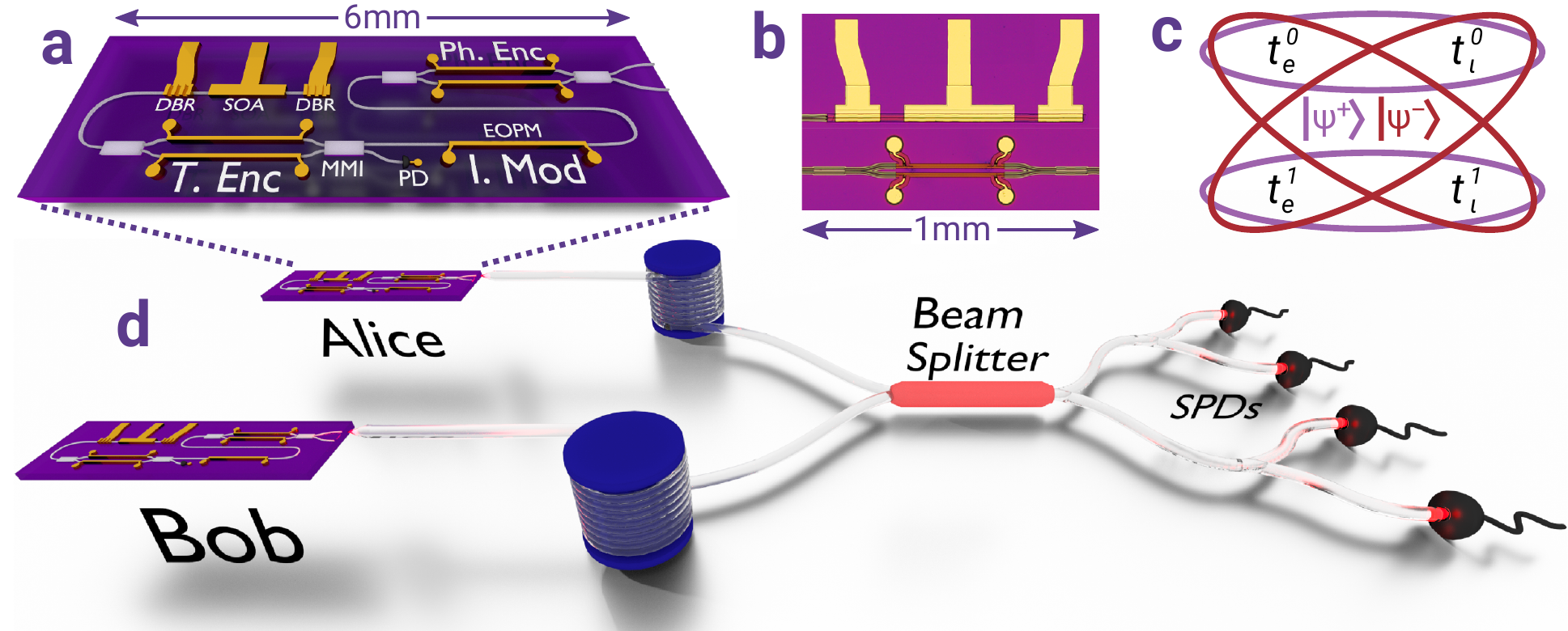}
	\caption{\textbf{Integrated MDI-QKD:} \textbf{a} Schematic of the $6\times2$~mm$^2$ InP chips used to generate the time-bin encoded BB84 weak coherent states through cascaded MZIs. Timing information is encoded with \textit{T.{ }Enc}, decoy intensities varied with \textit{I.{ }Mod} and phase encoded with \textit{Ph.{ }Enc}. A $1$~GHz photodiode (PD) can be used for power monitoring and feedback. \textbf{b} Microscope images of the on-chip DBR laser (top) and an MZI (bottom) measuring only $1$~mm in length. \textbf{c} Bell state projections for $\ket{\psi^+}$ (purple) and $\ket{\psi^-}$ (red) in time-bin encoding. $t^i_j$ corresponds to a detection event in the $i$th detector, where $j$ is either an early ($e$) or late ($l$) time-bin. \textbf{d} MDI-QKD experimental schematic. Two identical devices generate BB84 states independently and send them to the receiver. The states are projected in the Bell basis using a beam splitter and single-photon detectors (SPDs). A bank of 4 detectors is used to increase count rate, coincidence probability and key rate.}
	\label{fig:chip-mdi}
\end{figure*}

Quantum technologies promise a paradigm shift compared to their classical counterparts that will undermine our current methods of secure communication \cite{shor1994}. It will soon become necessary to deploy key exchange systems that are immune to such increases in computing power. Quantum key distribution (QKD) is one such approach which exploits quantum phenomena to exchange secret keys between distant parties without relying on assumed computationally hard problems \cite{BB84, ekert1991}. However, the stringent requirements for precise control has predominately limited QKD systems to small networks and laboratories. To realise ubiquitous quantum devices, new platforms are required for robust operation in harsh environments. 

Integrated photonics has seen vast improvements in recent years and represents a promising platform for mass-adoption of quantum technologies \cite{thompson2011}. In particular, indium phosphide (InP) offers crucial benefits for communication in a robust, phase-stable and compact platform. Lasers can be monolithically integrated with mW powers and narrow linewidths, fast electro-optic phase modulation can reach bandwidths of $40$~GHz and low-loss waveguides allow efficient routing \cite{Meint2014}. Such components mean that it is well suited for quantum communication protocols \cite{Sibson2017}. 

Quantum key distribution has been a leading quantum technology since its advent \cite{BB84, ekert1991} and has seen many proof-of-principle demonstrations, networks and commercial systems \cite{yin2016, Rubenok2011, comandar2016, zhang2018, commercial}. However, implementation security of these systems is an active area of research due to potential information leakage that is not considered in security proofs. Such side-channels may allow an eavesdropper to gain sensitive information during a key exchange \cite{Lo2014} or an attacker to manipulate a system and determine the secret key through classical means \cite{Lydersen2010}. 

To counter these attacks from a malicious adversary through uncharacterised side-channels, device-independent QKD schemes have been developed to limit the number of assumptions required for security \cite{masanes2011}. One such vulnerability is with single-photon detectors, for which measurement-device-independent quantum key distribution (MDI-QKD) has been proposed. This approach removes all possible attacks against the detection system~\cite{Lo2012}.

In this paper, we experimentally demonstrate MDI-QKD using cost-effective, mass-manufacturable, chip-based transmitters that could facilitate commercial quantum-secured communication. We show that $1$~kbps of secret key can be exchanged at $100$~km and predict positive key rates at more than $350$~km. The system removes detector vulnerabilities and represents a viable solution for near-term metropolitan quantum networks.

\section{Results}

\subsection{Protocol}

MDI-QKD removes all potential side channels on the detection system which could be exploited by a malicious adversary \cite{Lo2012}. A schematic of the experiment is shown in figure~\ref{fig:chip-mdi}. Unlike traditional point-to-point protocols, Alice and Bob act symmetrically by sending BB84 states to a third party, Charlie. Upon receipt of the states, Charlie measures the states in the Bell basis and publicly announces all successful events. The outcomes indicate quantum correlations between states but, without encoding knowledge known only by Alice and Bob, reveal no information about the secret key. This allows Charlie to be completely untrusted and it could even be assumed that an adversary is operating the receiver without compromising the security. By sharing the basis information for each state, Alice and Bob are able to infer a secret key which can be used in a symmetric key algorithm. 

As we use a weak coherent source we need to estimate the number of single-photon events. We employ a four-intensity decoy state analysis \cite{zhou2016} to bound the single-photon errors and yields. In this protocol, the $Z$ basis is used to generate key while the $X$ basis bounds the knowledge of an eavesdropper. 

While MDI-QKD typically offers a lower key rate at short distances when compared with point-to-point systems \cite{Sibson2017}, it can generate key rate at greater distances \cite{yin2016} as the errors are proportional to the square of the dark count probability. It also offers the potential for the measurement equipment to be shared between multiple parties through optical switching without compromising security.

\subsection{Transmitters}

Indium phosphide allows monolithic integration of all the required optical components for photonic quantum technology in a single platform \cite{Meint2014,Sibson2017}. Entirely on-chip components were used to generate high-fidelity, phase randomised, $250$~MHz clocked BB84 weak coherent states as required for quantum key distribution. A schematic of the chip is shown in figure~\ref{fig:chip-mdi}a. 

On-chip tunable distributed Bragg reflectors (DBRs) form an optical cavity around a waveguide integrated semiconductor optical amplifier (SOA) to create a Fabry-P\'{e}rot laser with $50$~dB sideband suppression and a $30$~pm linewidth. The linewidth measurement was limited by the optical spectrum analyser resolution ($30$~pm) and is expected to be smaller. The laser has a broad tuning capability of $10$~nm within the telecomms C-band through current injection of the DBRs. Alternatively, current injection of the SOA allowed fine tuning in steps of $80$~fm. Crucially, no wavelength filtering beyond the on-chip cavity was required ensuring easy wavelength tuning that is compatible with modern wavelength division multiplexing techniques while maintaining high-fidelity interference between independent devices.

Phase randomisation was achieved through gain switching of the SOA, as required by decoy state analysis \cite{zhou2016}. By exploiting the short upper-state and cavity lifetimes of the integrated laser, $-1.5$~V $200$~ps FWHM electrical pulses drain the cavity to generate $4$~ns phase randomised windows. At the start of a window the laser shows relaxation oscillations for $1$~ns until returning to a useful continuous operation. Quantum states can then be encoded in phase coherent time-bins.

\begin{figure}
	\centering
	\includegraphics[width = \linewidth]{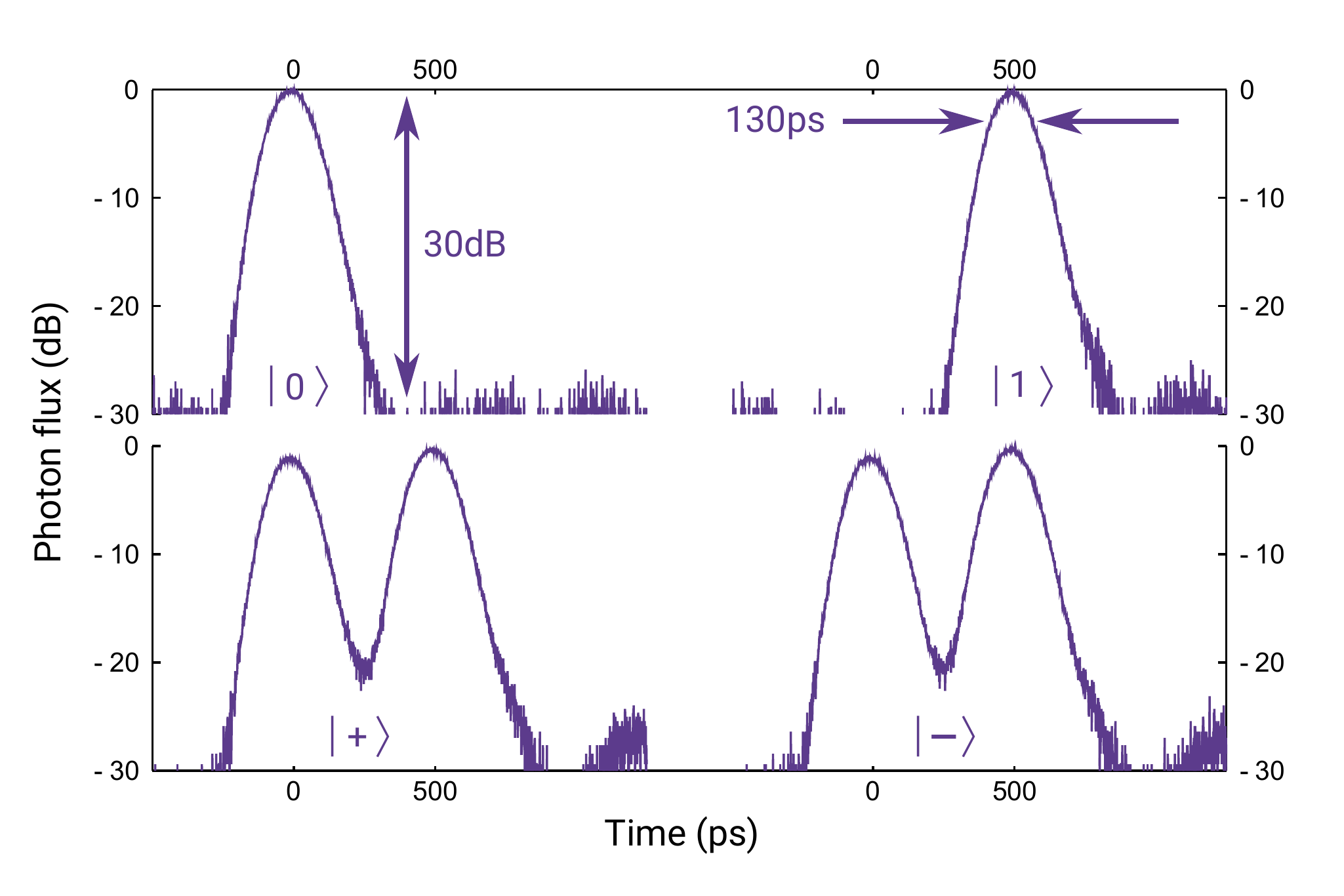}
	\caption{\textbf{BB84 States:} Histogram of the weak coherent states generated on-chip as measured by single-photon detectors. The time-bins are separated by $500$~ps and exhibit a $30$~dB extinction ratio and $130$~ps FWHM. }
	\label{fig:states}
\end{figure}

Electro-optic phase modulation (EOPM) was achieved through a quantum-confined Stark effect (QCSE) which has a bandwidth of $>10$~GHz. The efficacy of the QCSE means that the required voltages are lower when compared with other technologies, such as lithium niobate, and can be as small as $0.5$~mm in length. Together with multimode interferometers (MMIs), we create Mach-Zehnder interferometers (MZIs) that demonstrate $30$~dB extinction. $120$~ps electrical pulses are used to encode timing information in $500$~ps separated bins, resulting in $130$~ps FWHM pulses when detector jitter is considered. 

State intensity was modulated through the absorption of the QCSE to generate varying photon numbers for decoy state analysis. Square electrical signals allowed an intensity swing of $20$~dB to generate signal and decoy states. Only this stage of the state preparation required a multi-level electrical signal. However, the signals were still limited to only three levels in order to reduce the power consumption compared to an arbitrary level signal. A vacuum state was encoded by not pulsing the time encoding MZI. 

Finally, a relative phase between the two time-bins was required for the $X$ basis. A $\ket{+}$ state is defined with no phase between the early and late time-bins, while a relative $\pi$ phase is required to encode $\ket{-}$. To avoid phase dependent losses associated with the QCSE, phase between time-bins was encoded using a further MZI. By applying a phase shift to the top EOPM in the MZI in the first time-bin, and the bottom EOPM in the second, a $\pi$ phase is applied between the two pulses without decreasing the intensity of the state. A histogram of the four BB84 states is shown in figure~\ref{fig:states} which demonstrates the $30$~dB extinction ratio and $130$~ps FWHM.

Electrical signals were generated by synchronised pulse pattern generators, arbitrary waveform generators and FPGAs. All signals are kept below $4$~$\text{V}_\text{pp}$ which demonstrates reduced power requirements compared to fibre-based systems. Moreover, an effort has been made to reduce the number of multi-level signals required as generating two level states is less demanding than completely arbitrary waveforms. Such considerations will be crucial for wide adoption of quantum-secured communication.

\subsection{Detection}

\begin{figure}
	\centering
	\includegraphics[width = \linewidth]{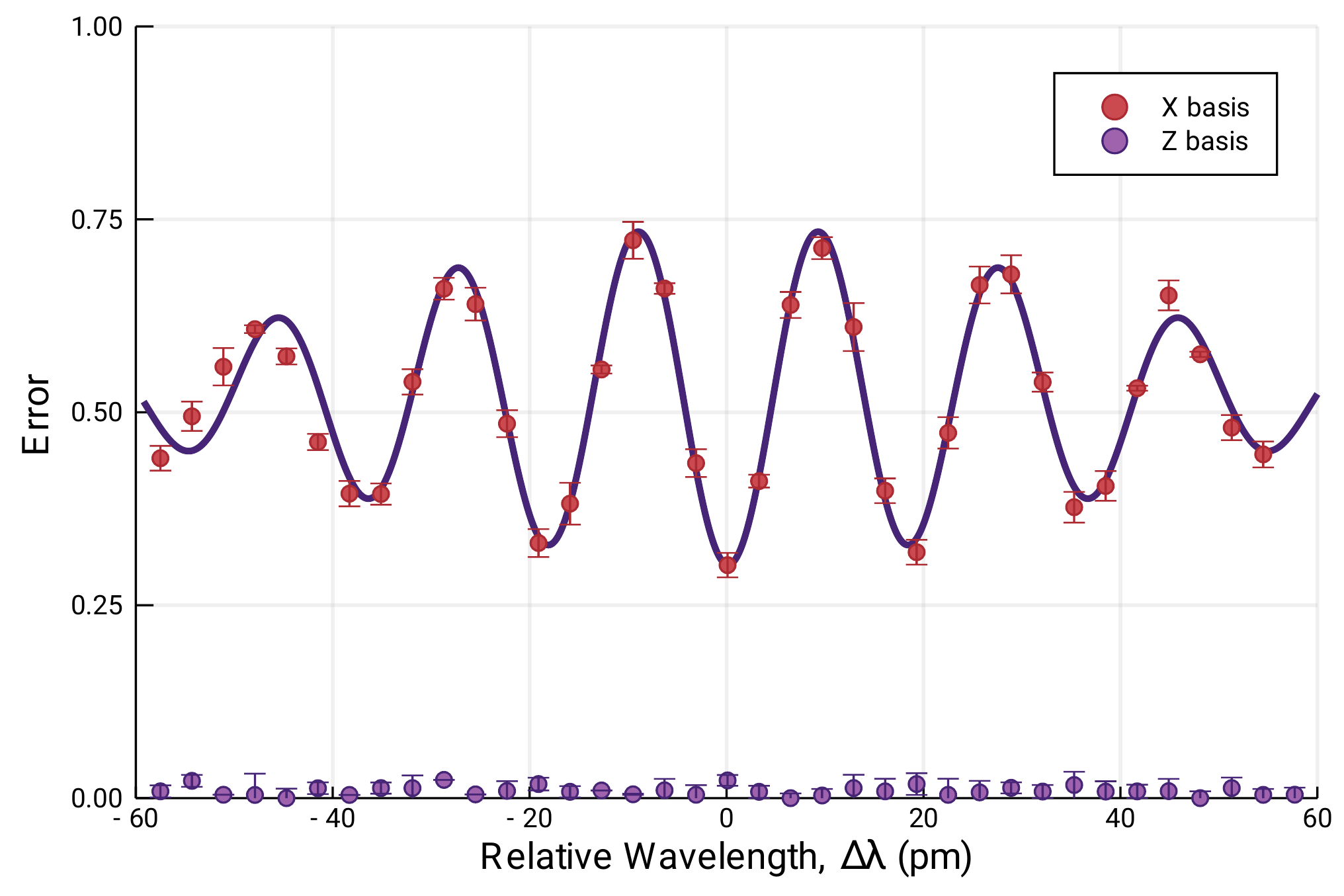}
	\caption{\textbf{Wavelength Overlap:} By varying the relative wavelength between the two lasers, the $X$ basis error oscillates as the time-bins tune in and out of phase. A reduction in visibility is due to the wavelength detuning. We demonstrate fine control to achieve a $30\%$ error, limited to a theoretical minimum of $25\%$. $Z$ basis errors are independent of the interference and remain below $1\%$.}
	\label{fig:bsm}
\end{figure}

Alice and Bob's states were projected in the Bell basis using a 50:50 fibre beam splitter and coincidences between single-photon detectors indicated successful measurements. In a time-bin encoding, coincidences between early and late bins differentiates between $\ket{\psi^\pm}$, as shown in figure~\ref{fig:chip-mdi}c. While is it only possible to project onto two of the four Bell states with linear optics \cite{Calsamiglia2001}, only a single Bell state is required for MDI-QKD \cite{Lo2012}. 

The Bell state projections require Hong-Ou-Mandel (HOM) interference \cite{HOM, Semenenko2019} at the beam splitter between states sent by Alice and Bob. Therefore, the two independent transmitters need to send states that are indistinguishable in time of arrival, polarisation, photon number and wavelength to ensure maximal interference.

The relative timing of the transmitters was controlled by tunable electronic delays with $1$~ps resolution and a classical, optical signal was used to synchronise the transmitter clocks with the time tagging electronics. Polarising beam splitters and polarisation controllers ensured the states sent by the transmitters were both linearly polarised and had a good overlap at the beam splitter. Photon number for each transmitter was calibrated independently and remained stable for the duration of the experiment.

The lasers were independently controlled and employed no feedback between the two devices during a key exchange to remain stable over several hours. Each transmitter wavelength was coarsely overlapped using an optical spectrum analyser. The lasers were then finely tuned in steps of $80$~fm through current injection of the SOA. Interference between the two devices was used to precisely overlap the wavelengths, as shown in figure~\ref{fig:bsm}. By using Alice's $\ket{+}$ and $\ket{-}$ states as a reference, Bob can vary the wavelength of his laser which changes the phase between his early and late time-bins. As the phase of Bob's qubits rotates, we find a sinusoidal beating where the visibility increases as the states become indistinguishable in wavelength. We demonstrate an error in the $X$ basis of $30\%$, which is limited to a theoretical minimum of $25\%$ due to the reduction in HOM interference from multiphoton terms \cite{Rarity2005}. Future systems could utilise this interference to provide active feedback to minimise errors between the transmitters for long-term stable operation. The $Z$ basis errors are not dependent on the interference and remained around $0.5\%$ during the sweep. 

In a time-bin encoding scheme only two detectors are required and $\ket{\psi^\pm}$ can be distinguished through coincident clicks between time-bins. However, while in principle $\ket{\psi^+}$ can be distinguished, due to the detector deadtime (typically $100$~ns) this event will never occur, reducing the number of successful events and key rate. While detectors exist with sub-nanosecond deadtime, they will typically sacrifice detection efficiency or wavelength tunability \cite{vetter2016,yun2019}. 

In this work, we utilise a banked detector system, which allows an increase in the number of detection events by allowing $\ket{\psi^+}$ to be detected with 50\% probability. This also means we can increase the number of detection events before detector saturation allowing higher key rates to be generated even at shorter distances. With waveguide integrated detectors being developed \cite{kahl2015}, we envisage systems with many detectors and integrated switches facilitating a completely integrated quantum key exchange.

The detectors used were superconducting nanowire single-photon detectors with an operating efficiency of $80\%$, dead time of $100$~ns, timing jitter of $30$~ps and dark count rates of $100$~Hz. Events were time tagged to give absolute timing of events from which coincidences and successful projections were determined. Bell states were then matched to corresponding states offline to determine errors and gains for key rate analysis.

\subsection{Rates}

\begin{figure}
	\centering
	\includegraphics[width = \linewidth]{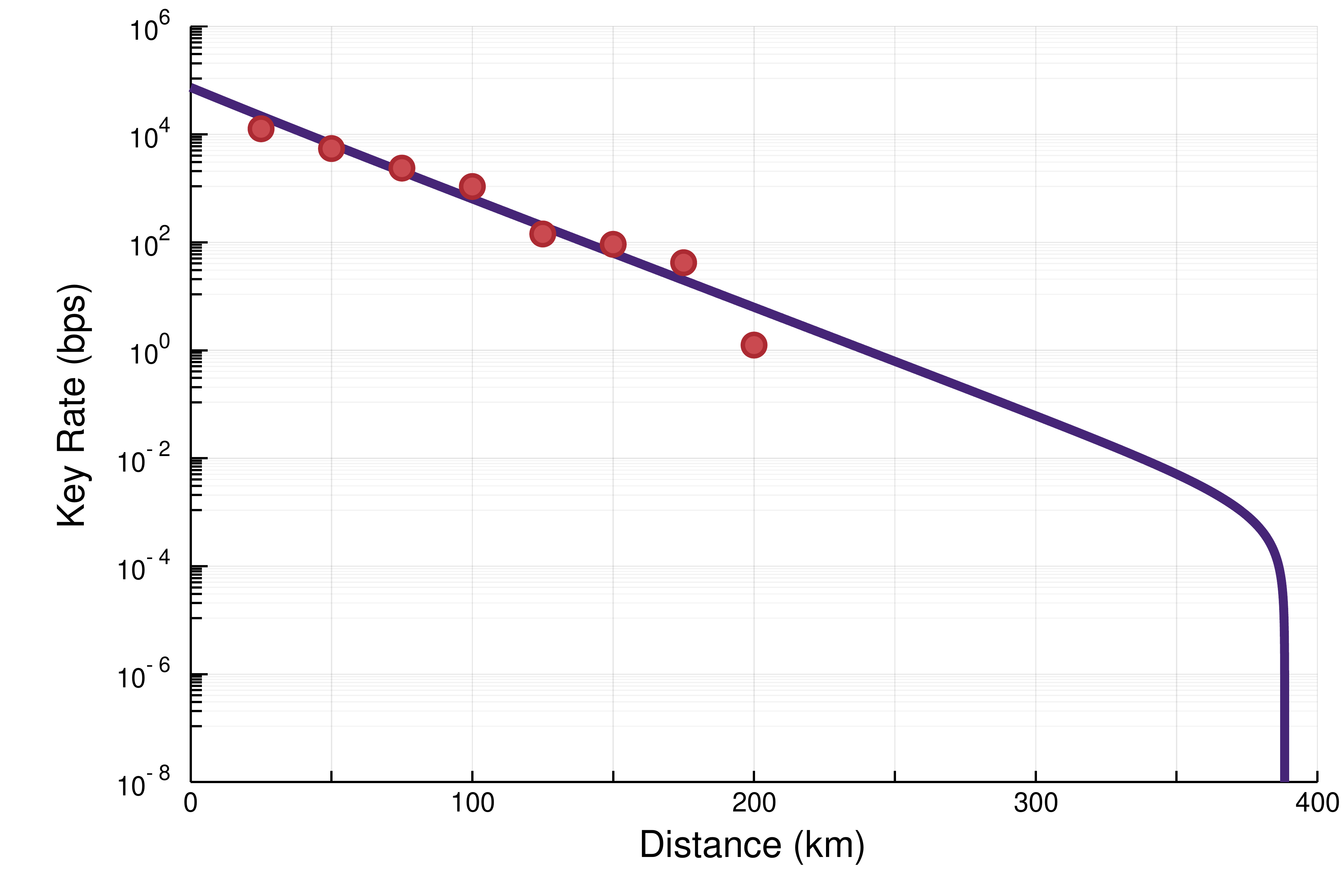}
	\caption{\textbf{Key rate:} Plot of the asymptotic key rates estimated from the system over an emulated fibre link assuming $0.2$~dB/km. We measure a secret key rate of $12$~kbps at $25$~km, while at $100$~km $1$~kbps of key can be securely exchanged. A model of the system using experimental parameters estimates that positive key rates are possible at distances of more than $350$~km.}
	\label{fig:key_rate}
\end{figure}

MDI-QKD has been demonstrated between two independent InP devices. Secret key rates were estimated over an emulated fibre link (using variable optical attenuators assuming $0.2$~dB/km) and are shown in figure~\ref{fig:key_rate}. At metropolitan distances ($25$~km), key rates of more than $12$~kbps are estimated at the asymptotic limit with positive key rates demonstrated up to $200$~km. Beyond this distance, the integration time required for a reasonable number detection events increases exponentially. For example, at $300$~km we would need to integrate for $6$ days. However, by characterising the experimental performance, we predict that a quantum-secured key exchange is possible at distances of more than $350$~km. 

We show that interference between independent transmitters is possible for $500$~ps separated ($2$~GHz clocked) time-bin encoded states with state of the art quantum bit error rates. We find an error of $30\%$ in the $X$ basis, which is limited to a theoretical minimum of $25\%$, demonstrating a good indistinguishability in all degrees of freedom. In the $Z$ basis, we achieve a quantum bit error rate of $0.5\%$. 

The mean photon number was $0.2$ in the $Z$ basis for the signal states. In the $X$ basis the mean photon numbers were $0.1$ and $0.01$ for decoy state analysis. The vacuum state intensities are kept at $5\times 10^{-4}$. Low photon numbers were used to limit the saturation of the detectors at low channel losses and allow positive key generation at further distances. The transmitter electronics is clocked at $2$~GHz with a state being sent every 8 clock cycles giving a $250$~MHz qubit rate. The bases were biased to produce an equal number of $Z$ and $X$ states, and therefore each of the $X$ decoy states were sent one third as often as a $Z$ signal state.

\section{Discussion}

In summary, we have demonstrated indium phosphide as a potential future platform for quantum-secured communication through MDI-QKD. Over an emulated fibre link, we show $1$~kbps of estimated secret key at $100$~km, positive key rates as far as $200$~km and estimate the system would provide positive key rates at distances of more than $350$~km. Quantum bit error rates below $0.5\%$ are demonstrated along with high-fidelity interference between independent devices at a qubit rate of $250$~MHz, comparable to previous demonstrations \cite{Rubenok2011, comandar2016, yin2016}.

Integrated photonics offers benefits for future networks with reduced power, weight and size requirements while simultaneously facilitating increased complexity with inherent phase stability. Indium phosphide devices are shown as a feasible platform for QKD networks, allowing relatively cost-effective devices to be easily mass manufactured. Integrated laser sources and efficient phase modulation satisfy all the requirements of high-fidelity quantum state preparation in a single monolithically fabricated platform. 

The topology of MDI-QKD means that citywide resource sharing can be achieved through commercially available optical switches at an untrusted centralised location. Furthermore, banks of detectors can be used to increase secret key rates. Advances in cryogenic cooling mean superconducting detectors are becoming more readily available and will likely be a vital part of future quantum-compatible networks. Such nodes will form the basis for more complex communication protocols that will require quantum repeaters and photonic information processing \cite{wehner2018}. 

The simplicity of the receiver in MDI-QKD lends itself towards an integrated platform \cite{Wang2019}. Waveguide integrated single-photon detectors \cite{sprengers2011}, on-chip wavelength demultiplexing \cite{sugita2000} and cryogenic optical switching \cite{eltes2019} mean that a completely integrated receiver device could further decrease the cost of QKD systems. Fully integrated measurement devices facilitate a drastic increase in the number of detectors to allow a higher count rate before saturation and relax the need for sub-nanosecond deadtimes. Key rates could be further increased through wavelength division multiplexing which can also allow coexistence with classical signals \cite{price2018}. Specialised electronics could be used to truly take advantage of the size, weight and power efficiency of an integrated system \cite{valivarthi2017}.

It is becoming increasingly vital that the future of secure communication is addressed to counter advances in classical and quantum computing. While quantum key distribution has been demonstrated as a potential candidate in future networks it has yet to be widely adopted. Concerns of side-channel attacks on physical implementations undermines the security promises of QKD systems. Here, we have improved on previous demonstrations of integrated QKD systems by removing all detection side-channels which vastly increases confidence in the security of the system. Mass-manufacturability and robust operation mean that integrated systems are poised to create an accessible platform for widespread quantum-secured communication.

\section*{Acknowledgements}

The authors thank Oclaro for the fabrication of devices through a PARADIGM project and A.~B. Price and G.~ D. Marshall for useful discussions. Funding was provided by the Engineering and Physical Sciences Research Council (EPSRC) (EP/L015730/1,  EP/M013472/1, EP/L024020/1, EP/N015126/1) and the European Research Council (ERC) (ERC-2014-STG 640079).

\bibliography{references}

\end{document}